\documentclass[useAMS,usenatbib]{mn2e}
\usepackage{epsfig,graphicx,lscape}
\title[Fe X lines in SERTS spectra]{Emission lines of Fe X in active region spectra obtained with the 
Solar Extreme-ultraviolet Research Telescope and Spectrograph}

\author[F. P. Keenan et al.]{F. P. Keenan\thanks{E-mail:
F.Keenan@qub.ac.uk},$^{1}$ D. B. Jess,$^{1}$ K. M. Aggarwal,$^{1}$
R. J. Thomas,$^{2}$ J. W. Brosius$^{2,3}$ \newauthor
and J. M. Davila$^{2}$
\\
$^{1}$Astrophysics Research Centre, School of Mathematics and Physics, Queen's University, Belfast BT7 1NN
\\
$^{2}$Laboratory for Solar Physics, Code 671, Heliophysics Science Division,
NASA's Goddard Space Flight Center, Greenbelt, 
\\
MD 20771, USA
\\
$^{3}$Department of Physics, The Catholic University of
America, Washington, DC 20064, USA
}

\begin{document}

\date{Received 2008 XXXX; in original form 2008 XXXX}

\pagerange{\pageref{firstpage}--\pageref{lastpage}} \pubyear{}

\maketitle

\label{firstpage}

\begin{abstract}
Fully relativistic calculations of radiative rates and electron impact
excitation cross sections for Fe\,{\sc x} are used to derive theoretical 
emission-line ratios involving transitions in the 174--366\,\AA\ 
wavelength range. A comparison of these with solar active region
observations obtained during the 1989 and 1995 flights of the 
Solar Extreme-ultraviolet Research Telescope
and Spectrograph (SERTS) reveals 
generally very good agreement between theory and
experiment. Several Fe\,{\sc x}
emission features are detected for the first time in SERTS spectra, while the 
3s$^{2}$3p$^{5}$ $^{2}$P$_{3/2}$--3s$^{2}$3p$^{4}$($^{1}$S)3d $^{2}$D$_{3/2}$ 
transition at 195.32\,\AA\ is identified for the first time (to our knowledge)
in an astronomical source. 
The most useful Fe\,{\sc x}
electron density (N$_{e}$)
diagnostic
line ratios are assessed to be 175.27/174.53 and 175.27/177.24, which both
involve lines close in wavelength and free from blends, 
vary by factors of 13 between N$_{e}$ = 10$^{8}$ and 10$^{11}$\,cm$^{-3}$, and 
yet show little
temperature sensitivity. Should these lines not be available, then the 257.25/345.74 ratio may be employed
to determine N$_{e}$, although this requires an accurate 
evaluation of the
instrument intensity calibration over a relatively large wavelength range.
However, if the weak 324.73\,\AA\ line of Fe\,{\sc x}
is reliably detected, the use of 324.73/345.74 or 257.25/324.73 
is recommended over 257.25/345.74.
Electron densities deduced from 175.27/174.53 and 175.27/177.24 
for the stars
Procyon and $\alpha$~Cen, using observations from the Extreme-Ultraviolet Explorer (EUVE) satellite,
are found to be consistent and in agreement with the values of N$_{e}$
determined from other diagnostic ratios in the EUVE spectra.  
A comparison of several theoretical extreme-ultraviolet Fe\,{\sc x} line ratios
with experimental values for a $\theta$-pinch, for which the plasma parameters have been
independently determined, reveals reasonable agreement between theory and observation,
providing some independent 
support for the accuracy of the adopted atomic data.

\end{abstract}


\section{Introduction}

Emission lines arising from transitions in Fe\,{\sc x}
have been widely detected in solar extreme-ultraviolet (EUV)
spectra (see, for example, Dere 1978; Thomas \& Neupert 1994).
Jordan (1965) first proposed the use of Fe\,{\sc x} lines
to determine the electron density in the solar corona,
while 
Jordan (1966) employed EUV transitions of Fe\,{\sc x}
and other Fe ions to derive both the electron density and coronal
Fe abundance. Since then, several authors have undertaken 
analyses of the solar EUV spectrum of Fe\,{\sc x}, including for example
Nussbaumer (1976) and Bhatia \& Doschek (1995).
To date, the most complete study is probably that of Del Zanna,
Berrington \& Mason (2004), which also provides
an excellent review of previous work on Fe\,{\sc x}.
The theoretical line ratios calculated by Del Zanna et al. employ radiative rates
generated with the {\sc superstructure} code by either Bhatia \& Doschek (1995) or 
themselves. For electron impact excitation 
rates, they use results for transitions among the lowest 31 fine-structure levels of Fe\,{\sc x}
calculated with the Breit-Pauli {\sc rmatrx} code, either by Pelan \& Berrington (2001) or
once again by themselves. 

Recently, Aggarwal \& Keenan (2004, 2005) have used the fully relativistic
{\sc grasp} and Dirac {\sc rmatrx} codes to calculate radiative rates and electron impact 
excitation cross sections, respectively, for transitions among the energetically
lowest 90 fine-structure levels of Fe\,{\sc x}. In this paper we use these results, 
plus additional atomic data presented here, 
to
generate theoretical emission-line ratios for Fe\,{\sc x}, and compare these
with high resolution spectra from the Solar Extreme-ultraviolet Research
Telescope and Spectrograph (SERTS).
Our aims are threefold, namely to (i) assess the importance of blending in the SERTS observations,
(ii) detect new Fe\,{\sc x} emission lines, and (iii) identify the best
Fe\,{\sc x} line ratios for use as electron density diagnostics.

This work is of particular relevance due to the recent launch
of the {\em Hinode} mission, which has on board the EUV Imaging
Spectrometer (EIS), covering the 170--211\,\AA\ and 246--292\,\AA\ wavelength ranges
(Culhane et al. 2007), similar to the SERTS spectral coverage of 170--225\,\AA\ and
235--450\,\AA\ (Thomas \& Neupert 1994).
It is clearly important that emission lines observed by
the EIS are fully assessed, and the best diagnostics identified.
SERTS provides the ideal testbed for this, due to its larger wavelength
coverage, allowing more lines from the same species to be detected and compared with theoretical
predictions. Furthermore, 
the best SERTS spectral resolution is about 0.03\,\AA\ [full width at half-maximum (FWHM)], obtained for the 
170--225\,\AA\ wavelength range observed in
second-order (Brosius, Davila \& Thomas 1998a),
which is a full factor of two better
than the 0.065--0.075\,\AA\ resolution available from EIS (Young et al. 2007).
Hence the SERTS 
data sets should allow emission features to be resolved and assessed which are 
blended in EIS spectra. 
Indeed, this is illustrated by our previous work on Fe\,{\sc xiii} (Keenan et al. 2007),
where we resolved the 203.79 and 203.83\,\AA\ features which are blended in EIS 
observations 
(Young et al. 2007).

\section[]{Observational data}

SERTS has had a total of 10 
successful flights over the period 1982--2000, 
each featuring a somewhat different set of optical components and observing targets.  We have selected 
data from two of these for analysis here, as they provide 
the most complete sequence of well-observed Fe\,{\sc x}
emission lines.

The flight on 1989 May 5 (henceforth referred to as SERTS--89) carried a standard gold-coated toroidal diffraction grating, and was the first to observe a strong solar active region, NOAA 5464.  
It detected hundreds of first-order emission lines in the 235--450\,\AA\ wavelength range, 
plus dozens of features spanning 170--225\,\AA, which appeared in second-order among the 340--450\,\AA\ first-order lines (Thomas \& Neupert 1994).  
The spectrum was recorded on Kodak 101--07 emulsion, at a spectral resolution of 50--80\,m\AA\ 
(FWHM) in first-order, and a spatial resolution of approximately 7 arcsec (FWHM).  This combination of high spectral resolution, strong signal, and broad wavelength coverage remains unsurpassed even 
today as the best available data set for investigating imaged solar emission features over the full 
wavelength range of 235--450\,\AA, and hence are used in the present
paper.  
Subsequent SERTS flights employed either a multilayer-coated diffraction grating or CCD-detector arrays which provided important technical advantages, but which also restricted the spectral bandpass that could be covered, making their data less suitable for the present study (see Keenan et al. 2007 and references therein).

The one exception was the flight on 1995 May 15 (henceforth referred to as SERTS--95) 
which incorporated a multilayer-coated toroidal diffraction grating that enhanced its sensitivity to second-order features in the 170--225\,\AA\ wavelength range (Brosius et al. 1998a).  
It observed active region NOAA 7870, again using Kodak 101--07 emulsion, 
and had an improved spatial resolution of approximately 5 arcsec.  Its unique shortwave multilayer coating led to the detection of many second-order emission lines not seen on previous SERTS flights (Thomas \& Neupert 1994; Brosius et al. 1996), and furthermore obtained the highest spectral resolution (0.03\,\AA\ FHWM) 
ever achieved for spatially resolved active region spectra in this wavelength range.  The SERTS--95 data therefore provide the best observations for investigating Fe\,{\sc x}
emission lines in the 170--225\,\AA\ region, and hence are employed in this paper.

Further details of the SERTS--95 observations, and the wavelength and absolute flux calibration procedures employed in the data reduction, may be found in Brosius et al. (1998a,b).  
Similar information for the SERTS--89 spectrum is available from Thomas \& Neupert (1994).  
We note that although the relative intensity 
calibration curves in both cases 
involved fitting to calculations of insensitive line-ratio pairs, Fe\,{\sc x}
lines made up less than 10\%\ of the points utilized in those procedures.
Furthermore, 
detailed comparisons of 
line ratio calculations for many ions with SERTS measurements 
have revealed
generally very good agreement between theory and observation,
providing support for the 
SERTS intensity calibrations (see, for example, Keenan et al. 2007 and references therein). 
Thus, even 
without fully independent intensity
calibrations, the SERTS results still provide a valid test for the new Fe\,{\sc x}
calculations presented here.  
It should also be noted that the SERTS--95 first-order calibration
curve was extrapolated from its fitted range of 238--336\,\AA\ to 
cover its complete bandpass. Hence there is some extra uncertainty in the
 intensities of lines
with wavelengths greater than 340\,\AA\ measured from this flight.  
However, no such problem exists with the SERTS--89 measurements of these lines.

We have searched for Fe\,{\sc x} emission lines in the SERTS--89 and SERTS--95 spectra using the
detections of Thomas \& Neupert (1994) and Brosius et al. (1998b), supplemented with those
from other sources, including
the NIST database,\footnote{http://physics.nist.gov/PhysRefData/} 
the latest version (V5.2) of the {\sc chianti}
database (Dere et al. 1997; Landi et al. 2006), the Atomic Line List of 
van Hoof,\footnote{http://www.pa.uky.edu/$\sim$peter/atomic/} and in particular the excellent
summary of line identifications by Del Zanna et al. (2004).
The latter provides not only a comprehensive list of wavelengths for 
well-observed Fe\,{\sc x} transitions, but also indicates 
alternative wavelengths where previous identifications are not consistent with their
conclusions. In Table 1 we list the Fe\,{\sc x}
transitions found in the SERTS--89 and SERTS--95 spectra, 
along with their measured
wavelengths. We also indicate possible blending features or alternative
identifications as suggested by Thomas \& Neupert or 
Brosius et al. in their original line lists.

\begin{table*}
 \centering
 \begin{minipage}{140mm}
  \caption{Fe\,{\sc x} line identifications in the SERTS 1989 and 1995 active region spectra.}
  \begin{tabular}{cll}
  \hline
Wavelength (\AA)    &   Transition  & Note$^{a}$
\\
\hline
174.53 & 3s$^{2}$3p$^{5}$ $^{2}$P$_{3/2}$--3s$^{2}$3p$^{4}$($^{3}$P)3d $^{2}$D$_{5/2}$ 
\\
175.27 & 3s$^{2}$3p$^{5}$ $^{2}$P$_{1/2}$--3s$^{2}$3p$^{4}$($^{3}$P)3d $^{2}$D$_{3/2}$ 
\\
175.48 & 3s$^{2}$3p$^{5}$ $^{2}$P$_{3/2}$--3s$^{2}$3p$^{4}$($^{3}$P)3d $^{2}$P$_{1/2}$ 
\\
177.24 & 3s$^{2}$3p$^{5}$ $^{2}$P$_{3/2}$--3s$^{2}$3p$^{4}$($^{3}$P)3d $^{2}$P$_{3/2}$ 
\\
180.38 & 3s$^{2}$3p$^{5}$ $^{2}$P$_{1/2}$--3s$^{2}$3p$^{4}$($^{3}$P)3d $^{2}$P$_{1/2}$ & Blended with Fe\,{\sc xi}
180.38 + Fe\,{\sc xvi} 360.76 (first-order). 
\\
184.53 & 3s$^{2}$3p$^{5}$ $^{2}$P$_{3/2}$--3s$^{2}$3p$^{4}$($^{1}$D)3d $^{2}$S$_{1/2}$ 
\\
190.05 & 3s$^{2}$3p$^{5}$ $^{2}$P$_{1/2}$--3s$^{2}$3p$^{4}$($^{1}$D)3d $^{2}$S$_{1/2}$ 
\\
193.72 & 3s$^{2}$3p$^{5}$ $^{2}$P$_{3/2}$--3s$^{2}$3p$^{4}$($^{1}$S)3d $^{2}$D$_{5/2}$ 
\\
195.32 & 3s$^{2}$3p$^{5}$ $^{2}$P$_{3/2}$--3s$^{2}$3p$^{4}$($^{1}$S)3d $^{2}$D$_{3/2}$ 
\\
201.56 & 3s$^{2}$3p$^{5}$ $^{2}$P$_{1/2}$--3s$^{2}$3p$^{4}$($^{1}$S)3d $^{2}$D$_{3/2}$ 
\\
220.26 & 3s$^{2}$3p$^{5}$ $^{2}$P$_{3/2}$--3s$^{2}$3p$^{4}$($^{3}$P)3d $^{2}$F$_{5/2}$ 
\\
256.43 & 3s$^{2}$3p$^{5}$ $^{2}$P$_{3/2}$--3s$^{2}$3p$^{4}$($^{3}$P)3d $^{4}$D$_{3/2}$ 
\\
257.25 & 3s$^{2}$3p$^{5}$ $^{2}$P$_{3/2}$--3s$^{2}$3p$^{4}$($^{3}$P)3d $^{4}$D$_{5/2,7/2}$ 
\\
324.73 & 3s$^{2}$3p$^{4}$($^{3}$P)3d $^{4}$D$_{7/2}$--3s3p$^{5}$($^{3}$P)3d $^{4}$F$_{9/2}$ 
\\
337.24 & 3s$^{2}$3p$^{4}$($^{3}$P)3d $^{2}$F$_{7/2}$--3s3p$^{5}$($^{3}$P)3d $^{2}$F$_{7/2}$ & Listed as Ar\,{\sc viii} by Thomas \& Neupert (1994). 
\\
345.74 & 3s$^{2}$3p$^{5}$ $^{2}$P$_{3/2}$--3s3p$^{6}$ $^{2}$S$_{1/2}$ 
\\
365.57 & 3s$^{2}$3p$^{5}$ $^{2}$P$_{1/2}$--3s3p$^{6}$ $^{2}$S$_{1/2}$ 
\\
\hline
\end{tabular}

$^{a}$From Brosius et al. (1998b) or Thomas \& Neupert (1994).  
\end{minipage} 
\end{table*}

Intensities and 
line widths (FWHM) of the Fe\,{\sc x} features are given in Tables 2 and
3 for the SERTS--89 and SERTS--95 active regions, respectively,
along with the associated 1$\sigma$ errors. These were measured
using modified versions of the Gaussian fitting routines employed by Thomas \& Neupert (1994),
as discussed by Keenan et al. (2007).
As a consequence, the
intensities, FWHM values and their uncertainties listed in Tables 2 and
3 are somewhat different from those originally
reported in Thomas \& Neupert and Brosius et al. (1998b). Also,
a uniform factor of 1.24 has been applied here to all
SERTS--89 intensities, reflecting a more recent re-evaluation of its 
absolute
calibration scale.
Even so, in all directly comparable cases, the resulting
line intensity values usually differ only slightly from those previously obtained. 
For the SERTS--95 data set, several of the stronger first-order Fe\,{\sc x} lines
could also be detected, and their measurements are therefore included in Table 3 along with the
second-order results. However, for the SERTS--89 spectrum only the first-order Fe\,{\sc x}
features could be reliably identified. 

\begin{table}
 \centering
  \caption{Fe\,{\sc x} line intensities and widths from the SERTS 1989 active region spectrum.}
  \begin{tabular}{ccc}
  \hline
Wavelength    &   Intensity & Line width 
\\
(\AA) & (erg\,cm$^{-2}$\,s$^{-1}$\,sr$^{-1}$) & (m\AA) 
\\
\hline
256.43 & 174.8 $\pm$ 62.5 & 94 $\pm$ 28 
\\
257.25 & 166.2 $\pm$ 38.3 & 86 $\pm$ 12 
\\
324.73 & 12.2 $\pm$ 3.2 & 54 $\pm$ 8 
\\
337.24 & 16.5 $\pm$ 7.9 & 73 $\pm$ 21 
\\
345.74 & 93.6 $\pm$ 11.9 & 98 $\pm$ 4 
\\
365.57 & 53.1 $\pm$ 7.3 & 113 $\pm$ 7 
\\
\hline
\end{tabular}
\end{table}

\begin{table}
 \centering
  \caption{Fe\,{\sc x} line intensities and widths from the SERTS 1995 active region spectrum.}
  \begin{tabular}{ccc}
  \hline
Wavelength    &   Intensity & Line width 
\\
(\AA) & (erg\,cm$^{-2}$\,s$^{-1}$\,sr$^{-1}$) & (m\AA) 
\\
\hline
174.53 & 655.4 $\pm$ 82.0 & 41 $\pm$ 3 
\\
175.27 & 133.5 $\pm$ 30.5 & 27 $\pm$ 5 
\\
175.48 & 79.4 $\pm$ 20.9 & 41 $\pm$ 9 
\\
177.24 & 335.2 $\pm$ 41.9 & 39 $\pm$ 3 
\\
180.38 & 3345.1 $\pm$ 373.0 & 48 $\pm$ 3 
\\
184.53 & 163.6 $\pm$ 19.6 & 33 $\pm$ 3 
\\
190.05 & 47.6 $\pm$ 8.4 & 41 $\pm$ 5 
\\
193.72 & 6.1 $\pm$ 2.9 & 16 $\pm$ 6 
\\
195.32 & 6.8 $\pm$ 2.4 & 25 $\pm$ 3 
\\
201.56 & 71.4 $\pm$ 14.0 & 78 $\pm$ 12 
\\
220.26 & 32.9 $\pm$ 21.0 & 67 $\pm$ 20 
\\
256.43 & 223.7 $\pm$ 68.0 & 62 $\pm$ 16 
\\
257.25 & 159.2 $\pm$ 48.5 & 44 $\pm$ 10 
\\
345.74 & 92.7 $\pm$ 37.1 & 50 $\pm$ 15 
\\
365.57 & 35.7 $\pm$ 16.6 & 53 $\pm$ 21 
\\
\hline
\end{tabular}
\end{table}

In Figs 1--5 we plot portions of the SERTS--89 and SERTS--95 spectra 
containing Fe\,{\sc x} transitions, focusing on
emission lines
which have not previously been identified
in SERTS data sets. We 
note that several of these features 
have
line intensities and widths comparable to the noise fluctuations.
In these instances, the reality of the feature was confirmed
by a visual inspection of the original SERTS film.
However, the lines are weak and clearly further
observations to strengthen these identifications
would be desirable. 

\begin{figure}
\epsfig{file=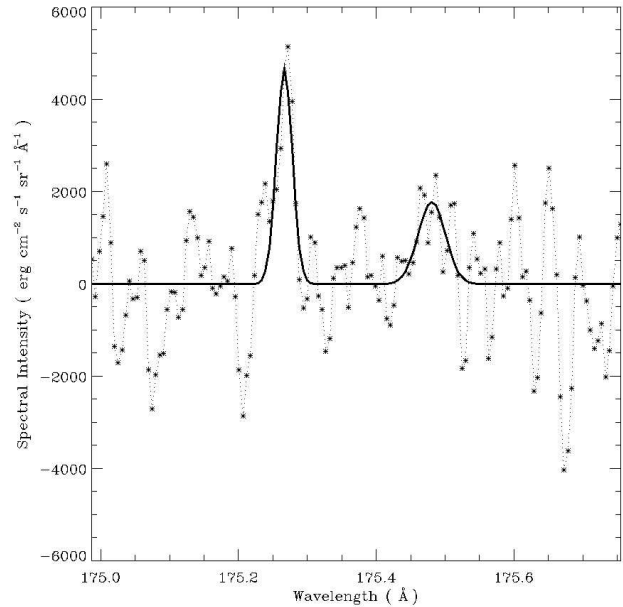,angle=0,width=8.5cm}
\caption{Plot of the SERTS 1995 active region spectrum in the
175.0--175.7\,\AA\ wavelength range.
The profile fit to the Fe\,{\sc x} 175.27 and Fe\,{\sc ix}/{\sc x} 
175.48\,\AA\ features
is shown by a solid line.} 
\end{figure}

\begin{figure}
\epsfig{file=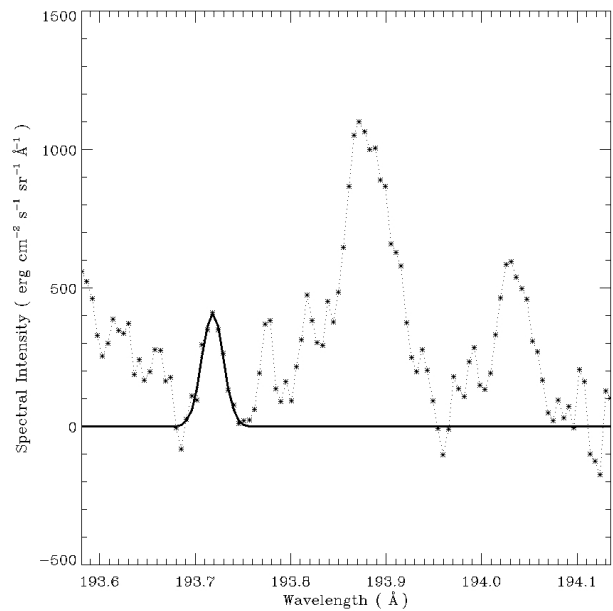,angle=0,width=8.5cm}
\caption{Plot of the SERTS 1995 active region spectrum in the
193.6--194.1\,\AA\ wavelength range.
The profile fit to the Fe\,{\sc x} 193.72\,\AA\ feature
is shown by a solid line. Also clearly visible in the figure are the 
Ca\,{\sc xiv} 193.87\,\AA\ and Ni\,{\sc xvi} 194.03\,\AA\ transitions.}
\end{figure}

\begin{figure}
\epsfig{file=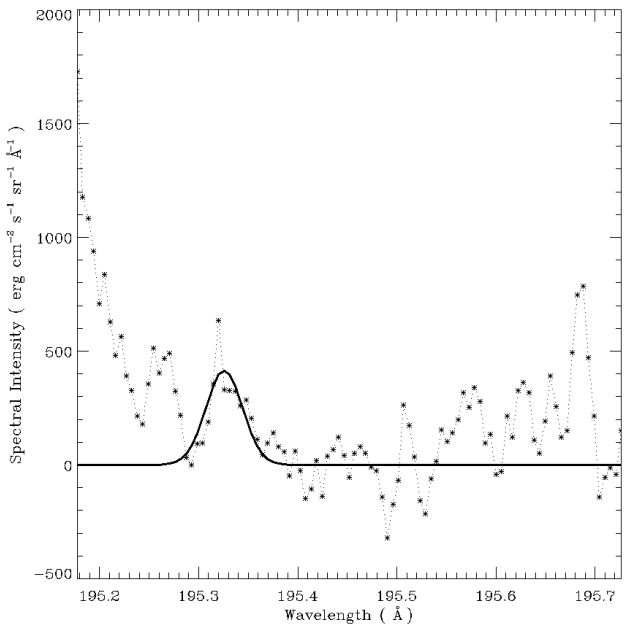,angle=0,width=8.5cm}
\caption{Plot of the SERTS 1995 active region spectrum in the
195.2--195.7\,\AA\ wavelength range.
The profile fit to the Fe\,{\sc x} 195.32\,\AA\ feature
is shown by a solid line. Also clearly visible is the wing of the
strong Fe\,{\sc xii} 195.12\,\AA\ transition.}
\end{figure}

\begin{figure}
\epsfig{file=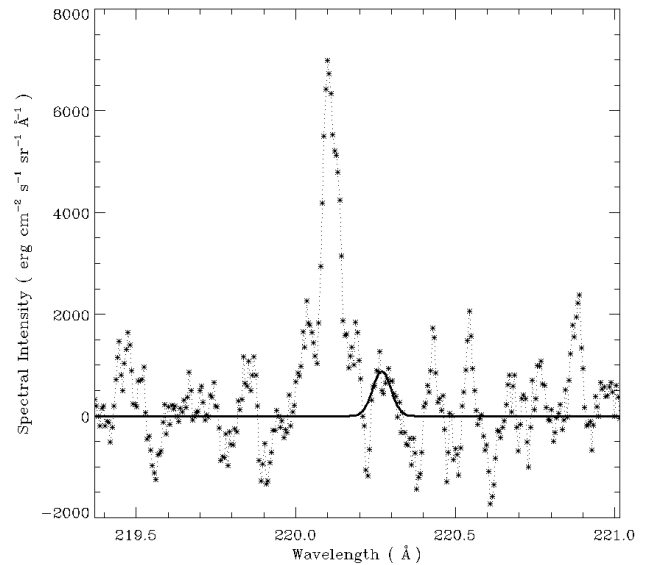,angle=0,width=8.5cm}
\caption{Plot of the SERTS 1995 active region spectrum in the
219.4--221.0\,\AA\ wavelength range.
The profile fit to the Fe\,{\sc x} 220.26\,\AA\ feature
is shown by a solid line. Also clearly visible in the figure is the 
Fe\,{\sc xiv} 220.09\,\AA\ transition.} 
\end{figure}

\begin{figure}
\epsfig{file=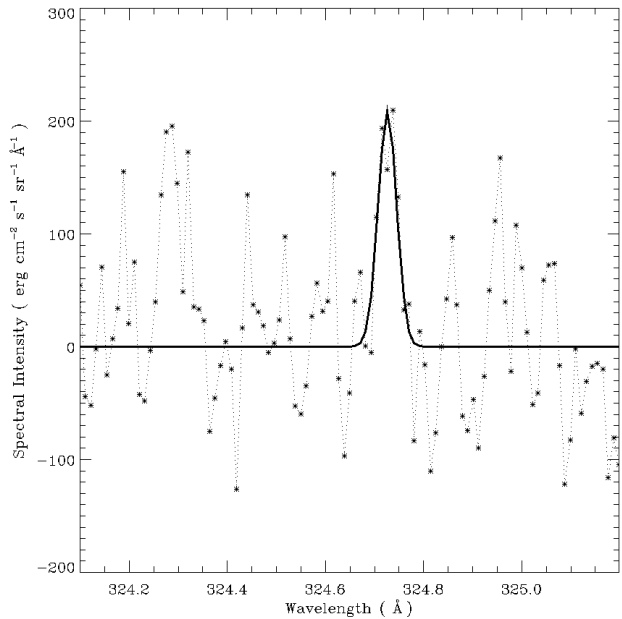,angle=0,width=8.5cm}
\caption{Plot of the SERTS 1989 active region spectrum in the
324.1--325.2\,\AA\ wavelength range.
The profile fit to the Fe\,{\sc x} 324.73\,\AA\ feature
is shown by a solid line.}
\end{figure}

\section{Theoretical line ratios}

The model ion for Fe\,{\sc x} consisted of the 90
energetically lowest fine-structure levels. These are listed in table 1 of 
Aggarwal \& Keenan (2005), and comprise levels arising from the 
3s$^{2}$3p$^{5}$, 3s3p$^{6}$, 3s$^{2}$3p$^{4}$3d, 
3s3p$^{5}$3d and 3s$^{2}$3p$^{3}$3d$^{2}$ configurations.
Where available, energies for these levels were taken from the
experimental compilation in the NIST database.
For the remainder, the theoretical values of Aggarwal \& Keenan 
were adopted. Test calculations including higher-lying levels, such as those
arising from the 
3s$^{2}$3p$^{4}$4s configuration, 
were found to have a negligible
effect on the theoretical line ratios
considered in this paper. 

The electron impact excitation cross sections employed in the present paper
are those calculated using the fully relativistic Dirac {\sc rmatrx} code by 
Aggarwal \& Keenan (2005).
For Einstein A-coefficients, Aggarwal \& Keenan (2004)
employed the fully relativistic {\sc grasp} code to generate
results for all 
transitions among the 54 
fine-structure levels of the 3s$^{2}$3p$^{5}$, 3s3p$^{6}$, 3s$^{2}$3p$^{4}$3d and 
3s3p$^{5}$3d configurations of Fe\,{\sc x}. 
Subsequently, Aggarwal \& Keenan (2005) extended this work to 
include the additional
36 levels 
arising from the 
3s$^{2}$3p$^{3}$3d$^{2}$ configuration,
but did not publish their results, and instead noted that
these were available from the authors on request. 
However, we provide these 
calculations in Table 4. 
A complete version of the
table is available in the electronic version of this paper, with only
sample results included in the hardcopy edition. The indices used to 
represent the lower and upper levels of a transition 
have been defined in table 1 of Aggarwal \& Keenan (2005). 
We note that radiative data 
for all 4005 transitions 
among the 90 levels considered in our 
calculations are available
in electronic form on request from one of the authors (K.Aggarwal@qub.ac.uk).
These results have been adopted for all the Fe\,{\sc x}
transitions, apart from 3s$^{2}$3p$^{5}$ $^{2}$P$_{3/2}$--3s$^{2}$3p$^{5}$ $^{2}$P$_{1/2}$, where we
have employed the A-value determined by Tr\"{a}bert, Saathoff \& Wolf (2004) based on the extrapolation of
experimental 
measurements for Co\,{\sc xi}, Ni\,{\sc xii} and Cu\,{\sc xiii}. However, we note that
this extrapolated result differs by less than 0.7 per cent from the calculation of
Aggarwal \& Keenan (2004).

\begin{table*}
 \centering
                                          
\caption{Sample of transition wavelengths ($\lambda_{ij}$ in \AA), radiative rates (A$_{ji}$ in s$^{-1}$), and oscillator strengths (f$_{ij}$, dimensionless), for electric       
dipole (E1), electric quadrupole (E2), magnetic dipole (M1), and magnetic quadrupole (M2) transitions in Fe\,{\sc x}.
($a{\pm}b \equiv a{\times}$10$^{{\pm}b}$). The full table is available as Supplementary Material to the
online article. }                
\begin{tabular}{rrrrrrrrrrr}                                                                                                                                              
\hline                                                                                                                                                                
$i$ & $j$ & $\lambda_{ij}$ & A$^{{\rm E1}}$  & f$^{{\rm E1}}$ & A$^{{\rm E2}}$  & f$^{{\rm E2}}$  & A$^{{\rm M1}}$  & f$^{{\rm M1}}$ & A$^{{\rm M2}}$  & f$^{{\rm M2}}$ \\

\hline                                                                                                                                                   
    1 &   55 &  1.228$+$02 &  0.000$+$00 &  0.000$+$00 &  3.327$+$04 &  1.127$-$07 &  3.222$-$02 &  1.092$-$13 & 
 0.000$+$00 &  0.000$+$00 \\         
    1 &   56 &  1.221$+$02 &  0.000$+$00 &  0.000$+$00 &  1.735$+$05 &  7.755$-$07 &  0.000$+$00 &  0.000$+$00 & 
 0.000$+$00 &  0.000$+$00 \\         
    1 &   57 &  1.211$+$02 &  0.000$+$00 &  0.000$+$00 &  4.263$+$02 &  9.370$-$10 &  4.064$-$03 &  8.932$-$15 & 
 0.000$+$00 &  0.000$+$00 \\         
    1 &   58 &  1.210$+$02 &  0.000$+$00 &  0.000$+$00 &  2.703$+$02 &  8.903$-$10 &  1.077$-$05 &  3.547$-$17 &
  0.000$+$00 &  0.000$+$00 \\         
    1 &   59 &  1.210$+$02 &  0.000$+$00 &  0.000$+$00 &  2.677$+$03 &  1.175$-$08 &  0.000$+$00 &  0.000$+$00 &
  0.000$+$00 &  0.000$+$00 \\         
    1 &   60 &  1.187$+$02 &  0.000$+$00 &  0.000$+$00 &  5.967$+$05 &  1.260$-$06 &  9.531$-$03 &  2.013$-$14 & 
 0.000$+$00 &  0.000$+$00 \\         
    1 &   61 &  1.184$+$02 &  0.000$+$00 &  0.000$+$00 &  7.096$+$05 &  2.239$-$06 &  7.619$-$04 &  2.404$-$15 & 
 0.000$+$00 &  0.000$+$00 \\         
    1 &   62 &  1.174$+$02 &  0.000$+$00 &  0.000$+$00 &  1.019$+$04 &  4.211$-$08 &  0.000$+$00 &  0.000$+$00 &
  0.000$+$00 &  0.000$+$00 \\         
    1 &   63 &  1.173$+$02 &  0.000$+$00 &  0.000$+$00 &  2.791$+$01 &  8.637$-$11 &  2.753$-$01 &  8.518$-$13 &
  0.000$+$00 &  0.000$+$00 \\         
    1 &   64 &  1.172$+$02 &  0.000$+$00 &  0.000$+$00 &  2.895$+$02 &  5.965$-$10 &  7.452$-$02 &  1.535$-$13 &
  0.000$+$00 &  0.000$+$00 \\         
    1 &   65 &  1.171$+$02 &  0.000$+$00 &  0.000$+$00 &  2.663$+$03 &  2.735$-$09 &  4.582$-$02 &  4.706$-$14 &
  0.000$+$00 &  0.000$+$00 \\

\hline
\end{tabular}
\end{table*}

Proton
impact excitation is generally only considered to be important for fine-structure transitions 
within the ground or first excited states of an ion (Seaton 1964),
and in the present paper we have employed the calculations of
Bely \& Faucher (1970) for 
3s$^{2}$3p$^{5}$ $^{2}$P$_{3/2}$--3s$^{2}$3p$^{5}$ $^{2}$P$_{1/2}$.
However, Bhatia \& Doschek (1995) have generated theoretical proton rates
for several of the 
3s$^{2}$3p$^{4}$($^{3}$P)3d $^{4}$D$_{J}$--3s$^{2}$3p$^{4}$($^{3}$P)3d $^{4}$F$_{J^{\prime}}$ 
and
3s$^{2}$3p$^{4}$($^{3}$P)3d $^{4}$D$_{J}$--3s$^{2}$3p$^{4}$($^{1}$D)3d $^{2}$P$_{1/2}$ 
transitions, and they have therefore been included for completeness.

Using the atomic data discussed above in conjunction
with a recently updated version of the statistical equilibrium
code of Dufton (1977), relative Fe\,{\sc x}
level populations and hence emission-line strengths were calculated
as a function of 
electron temperature (T$_{e}$) and density (N$_{e}$). Details of the
procedures involved and approximations made may be found in Dufton (1977) and
Dufton et al. (1978).
Results have been generated for 
a range of electron temperatures
(T$_{e}$ = 10$^{5.5}$--10$^{6.4}$\,K in steps of 0.1 dex) over which
Fe\,{\sc x}
has a fractional abundance in ionization equilibrium of 
N(Fe\,{\sc x})/N(Fe) $\geq$ 10$^{-4}$ (Bryans et al. 2006),
and for N$_{e}$ = 10$^{8}$--10$^{13}$\,cm$^{-3}$
in steps of 0.1 dex, which should be 
appropriate to most coronal-type plasmas.
These data 
are far too extensive to reproduce here, as with 90 fine-structure levels in our 
calculations we have intensities for 4005 transitions at each of the 
510 possible (T$_{e}$, N$_{e}$) combinations considered, yielding a total of
over 2 million data points. 
However, results 
involving any line pair, in either photon or energy units, are freely available from
one of the authors (FPK) by email on request.
Given expected errors in the adopted atomic data of typically less than
$\pm$20 per cent (see the references above), we estimate
the
theoretical ratios to be accurate to better than $\pm$30 per cent.

In Figs 6--10 we plot some sample theoretical 
emission-line ratios at the temperature of maximum Fe\,{\sc x}
fractional abundance in ionisation equilibrium, T$_{e}$ = 10$^{6}$\,K (Bryans et al. 2006), plus
$\pm$0.2 dex about this value. The transitions corresponding to the wavelengths
given in the figures are listed in Table 1.
An inspection of the figures reveals that the 175.27/174.53 and 175.27/177.24  
ratios
provide potentially excellent N$_{e}$--diagnostics, as they involve line pairs which are close in wavelength, 
vary by factors of 
13.1 and 13.4, respectively, between N$_{e}$ = 10$^{8}$ and 10$^{11}$\,cm$^{-3}$, and
show little T$_{e}$--sensitivity. 
For example, changing T$_{e}$ from 10$^{6}$ to 10$^{6.2}$\,K leads to a less than 5 per cent
variation in both ratios at N$_{e}$ = 10$^{10}$\,cm$^{-3}$, which in turn would result
in a less than 0.1 dex difference in the derived
value of N$_{e}$. The 
usefulness of these and other ratios as N$_{e}$--diagnostics is discussed in more detail
in Section 4.1.

\begin{figure}
\epsfig{file=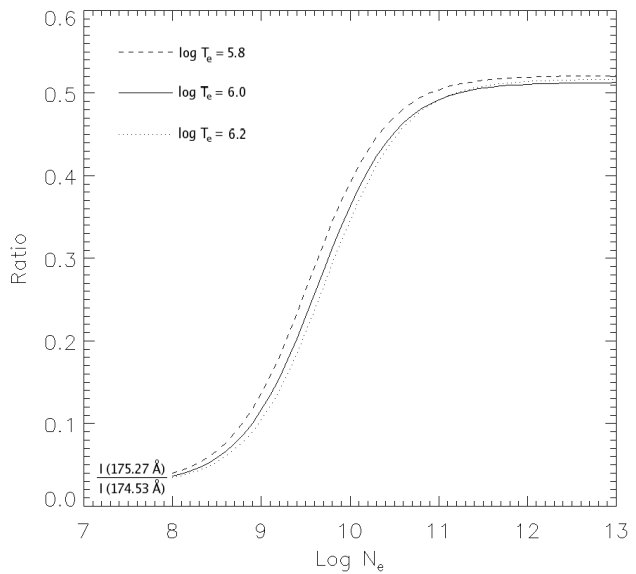,angle=0,width=8.5cm}
\caption{The theoretical Fe\,{\sc x}
emission-line intensity ratio
I(175.27\,\AA)/I(174.53\,\AA), where I is in energy units,
plotted as a function of logarithmic electron density
(N$_{e}$ in cm$^{-3}$) at the temperature
of maximum Fe\,{\sc x} fractional abundance in ionization
equilibrium, T$_{e}$ = 10$^{6}$\,K (Bryans et al.
2006), plus $\pm$0.2 dex about this
value. }
\end{figure}

\begin{figure}
\epsfig{file=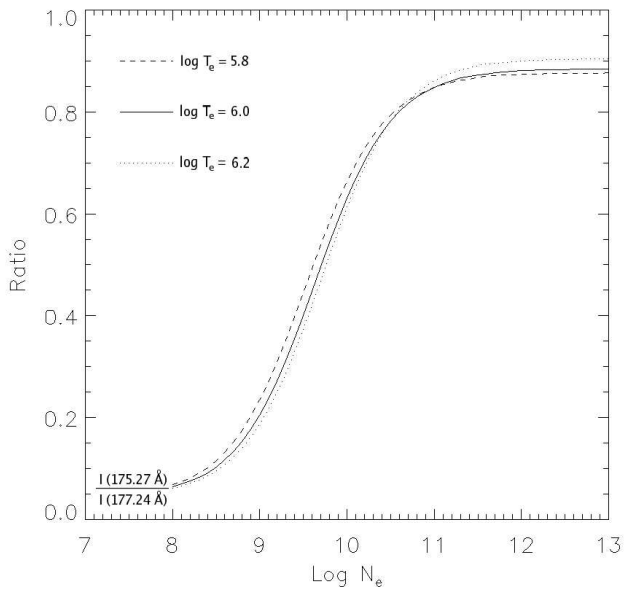,angle=0,width=8.5cm}
\caption{Same as Fig. 6 except for the 
I(175.27\,\AA)/I(177.24\,\AA) intensity ratio.}
\end{figure}

\begin{figure}
\epsfig{file=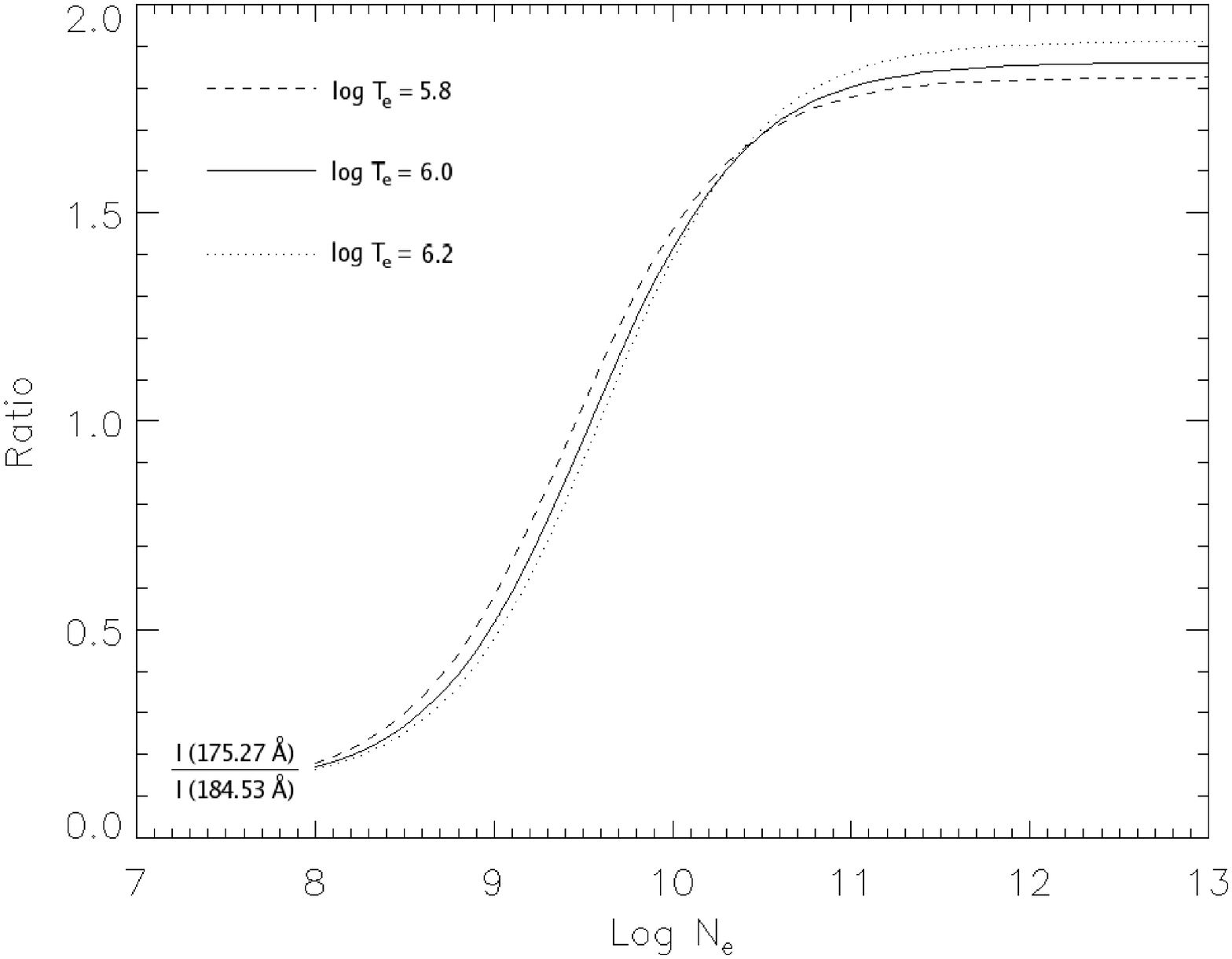,angle=0,width=8.5cm}
\caption{Same as Fig. 6 except for the 
I(175.27\,\AA)/I(184.53\,\AA) intensity ratio.}
\end{figure}

\begin{figure}
\epsfig{file=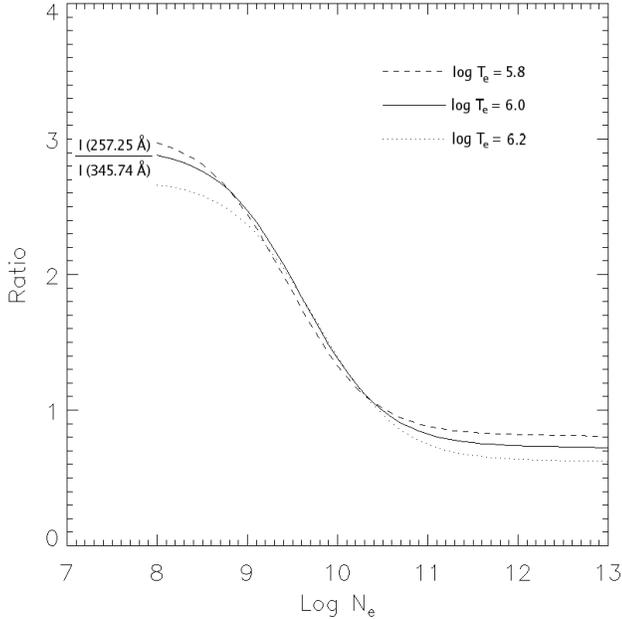,angle=0,width=8.5cm}
\caption{Same as Fig. 6 except for the 
I(257.25\,\AA)/I(345.74\,\AA) intensity ratio.}
\end{figure}

\begin{figure}
\epsfig{file=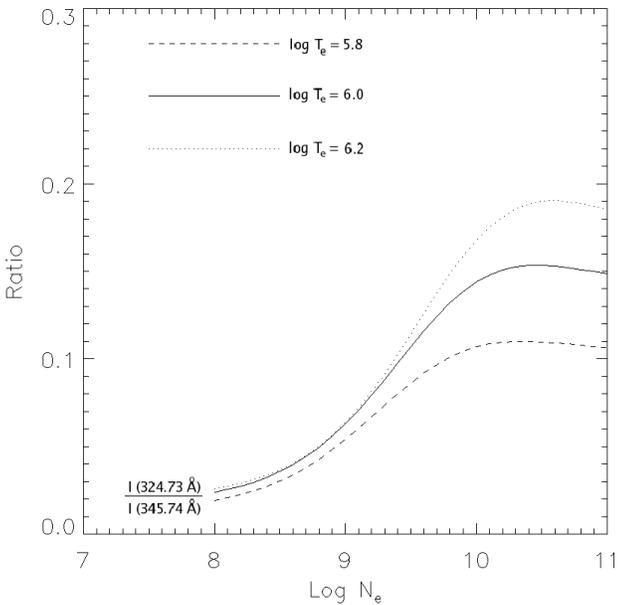,angle=0,width=8.5cm}
\caption{Same as Fig. 6 except for the 
I(324.73\,\AA)/I(345.74\,\AA) intensity ratio.}
\end{figure}

\section{Results and discussion}

Following Keenan et al. (2007), we categorise emission-line ratios as three 
types, namely:

(i) branching ratios (i.e. where the transitions arise from a common upper level), which
are predicted to be constant under normal plasma conditions. (An exception is where there is 
significant opacity, 
which does not apply to the Fe\,{\sc x}
lines considered here);

(ii) ratios which are predicted to be relatively insensitive to variations
in T$_{e}$ and N$_{e}$ over the range of plasma parameters of interest;

(iii)
ratios which are predicted to be strongly N$_{e}$--sensitive, and hence potentially
provide useful electron density 
diagnostics.

Clearly, ratios which fall into categories (i) or (ii) are the most useful for 
identifying and assessing the importance of blends, as well as investigating possible
errors or problems with the adopted
atomic data, 
as one does not need to reliably know the plasma electron temperature and density 
to calculate a line ratio for comparison with the observed value.
In Tables 5 and 6 we therefore list the 
observed line ratios for the SERTS--89 and SERTS--95
active regions, respectively (along with the associated 1$\sigma$ errors), which fall into
category (i) or (ii), along with theoretical values both from the 
present calculations
and the {\sc chianti} database, which employs the atomic data of Del Zanna et al. (2004).
Once again, following Keenan et al. (2007), for category (ii) ratios we have defined 
`relatively insensitive' as being those which
are predicted to vary by less than $\pm$20 per cent when the electron density 
is changed by a factor of 2 (i.e. $\pm$0.3 dex). As noted by Keenan et al., 
most of the electron densities derived for 
the SERTS--89 and
SERTS--95 active regions from species formed 
at similar temperatures to Fe~{\sc x}
are consistent with log N$_{e}$ = 9.4$\pm$0.3 for both solar features.
The theoretical results in Tables 5 and 6 have therefore been calculated at the temperature of
maximum Fe\,{\sc x} fractional abundance in ionization equilibrium, 
T$_{e}$ = 10$^{6}$\,K (Bryans et al. 2006), and for N$_{e}$ = 10$^{9.4}$\,cm$^{-3}$. 
However, we note that
changing the 
adopted value of T$_{e}$ 
by $\pm$0.2 dex or the density by $\pm$0.5 dex does not significantly alter 
the discussions below. 

\begin{table}
 \centering
\begin{minipage}{80mm}
  \caption{Comparison of theory and observation for emission-line intensity
ratios from the SERTS 1989 active region 
spectrum.}
  \begin{tabular}{cccc}
  \hline
Line ratio &   Observed & Present & {\sc chianti} 
\\
& & theory & theory
\\
\hline
\multicolumn{4}{l}{(i) Ratios of lines with common upper levels:}
\\[5pt]
365.57/345.74 & 0.57 $\pm$ 0.11 & 0.41 & 0.42
\\[10pt]
\multicolumn{4}{l}{(ii) Ratios which are only weakly N$_{e}$--dependent:$^{a}$}
\\[5pt]
256.43/345.74 & 1.9 $\pm$ 0.7 & 0.26 & 0.29
\\
324.73/257.25 & 0.073 $\pm$ 0.026 & 0.047 & 0.055 
\\
337.24/345.74 & 0.18 $\pm$ 0.09 & 0.11 & 0.22 
\\
\hline
\end{tabular}

$^{a}$Present theoretical ratios and those from {\sc chianti} calculated at T$_{e}$ = 10$^{6}$\,K and
N$_{e}$ = 10$^{9.4}$\,cm$^{-3}$.
\end{minipage} 
\end{table}

\begin{table}
 \centering
\begin{minipage}{80mm}
  \caption{Comparison of theory and observation for emission-line intensity
ratios from the SERTS 1995 active region 
spectrum.}
  \begin{tabular}{cccc}
  \hline
Line ratio &   Observed & Present & {\sc chianti} 
\\
& & theory & theory
\\
\hline
\multicolumn{4}{l}{(i) Ratios of lines with common upper levels:}
\\[5pt]
180.38/175.48 & 42 $\pm$ 12 & 2.7 & 3.1
\\
190.05/184.53 & 0.29 $\pm$ 0.06 & 0.32 & 0.28 
\\
201.56/195.32 & 11 $\pm$ 4 & 0.31 & 0.37
\\
365.57/345.74 & 0.39 $\pm$ 0.24 & 0.41 & 0.42
\\[10pt]
\multicolumn{4}{l}{(ii) Ratios which are only weakly N$_{e}$--dependent:$^{a}$}
\\[5pt]
174.53/177.24 & 2.0 $\pm$ 0.3 & 1.8 & 1.8
\\
175.48/177.24 & 0.24 $\pm$ 0.07 & 0.065 & 0.060
\\
180.38/177.24 & 10 $\pm$ 2 & 0.18 & 0.18
\\
184.53/177.24 & 0.49 $\pm$ 0.08 & 0.41 & 0.42
\\
190.05/177.24 & 0.14 $\pm$ 0.03 & 0.13 & 0.12
\\
193.72/177.24 & 0.018 $\pm$ 0.009 & 0.026 & 0.027
\\
195.32/177.24 & 0.020 $\pm$ 0.008 & 0.020 & 0.015
\\
201.56/177.24 & 0.21 $\pm$ 0.05 & 0.0064 & 0.0054
\\
220.26/177.24 & 0.098 $\pm$ 0.064 & 0.065 & 0.050
\\
256.43/345.74 & 2.4 $\pm$ 1.2 & 0.26 & 0.29
\\
\hline
\end{tabular}

$^{a}$Present theoretical ratios and those from {\sc chianti} calculated at T$_{e}$ = 10$^{6}$\,K and
N$_{e}$ = 10$^{9.4}$\,cm$^{-3}$.
\end{minipage} 
\end{table}

A comparison of the observed and theoretical 365.57/345.74 
ratios in Table 5 reveals that the former is 
slightly larger
than expected, suggesting that the 
365.57\,\AA\ transition
may be subject to a small amount of blending.
To investigate this, we have
generated a synthetic active region spectrum
using {\sc chianti}, which indicates blending
of the Fe\,{\sc x}
feature with Ne\,{\sc v} 365.60\,\AA. The latter is
predicted by {\sc chianti} to contribute about 50 per cent to the total
line intensity, although this estimate must be treated with caution. 
For example, 
{\sc chianti}
adopts solar photospheric abundances, which may not be applicable.
However, there are no blending
species identified by {\sc chianti} for the 345.74\,\AA\ line. 
The blending of 
365.57\,\AA\
was also found by Keenan \& Aggarwal (1989)
in their study of Ne\,{\sc v} 
lines in spectra from the S082A instrument on the 
{\em Skylab} mission.
On the other hand, the 345.74 and 365.57\,\AA\ lines are Fe\,{\sc x}
features which can be reliably measured in several SERTS 
data sets, as they are relatively strong, including SERTS--95 (see Table 6).
The average 365.57/345.74 ratio for a total of 4 SERTS spectra is
0.41 $\pm$ 0.12, in excellent agreement 
with the theoretical value of 0.41, implying that the blending of 365.57\,\AA\
with Ne\,{\sc v} 365.60\,\AA\ may not always be significant. 

The good agreement between observation and both the present and {\sc chianti}
line ratio calculations for 174.53/177.24, 190.05/177.24 and 324.73/257.25 
in Tables 5 and 6 indicate that these lines are well detected 
in the SERTS spectra and are free from blends. 
In particular, although {\sc chianti} lists a line of Fe\,{\sc xii}
at 190.07\,\AA\ that is predicted to be about 17 per cent
 of the intensity of
Fe\,{\sc x} 190.05\,\AA, 
the resolution of the SERTS--95 active region spectrum is such that this
feature appears as an asymmetry on the wing of the Fe\,{\sc x}
profile. Hence it could be modelled and removed from the total line intensity
measurement.
Furthermore, our identification of the 
3s$^{2}$3p$^{4}$($^{3}$P)3d $^{4}$D$_{7/2}$--3s3p$^{5}$($^{3}$P)3d $^{4}$F$_{9/2}$ line at 
324.73\,\AA\ is the first detection of this Fe\,{\sc x} feature in the SERTS spectrum.
Del Zanna et al. (2004) tentatively identified this transition as being a line at 
327.8\,\AA\ in the SOHO/CDS quiet Sun 
spectrum of Brooks et al. (1999). However,
we find no line at 327.8\,\AA, and our wavelength is in agreement with that
measured by Bhatia \& Doschek (1995) from spectra obtained
with the S082A instrument on {\em Skylab}, as well as the laboratory spectrum determination by Smitt (1977).

The experimental 190.05/184.53 and 184.53/177.24 ratios are both in good agreement 
with theory, indicating that the 184.53\,\AA\ line is reliably detected, although
the {\sc chianti} synthetic spectrum implies a small contribution (4 per cent) from 
an Ar\,{\sc xi}
transition.
By contrast, measured ratios with the 175.48, 180.38 or 
256.43\,\AA\ lines in the numerator are all much larger than the theoretical
values, due to strong blending with Fe\,{\sc ix} 175.48\,\AA, Fe\,{\sc xi} 180.38 plus
Fe\,{\sc xvi} 360.76\,\AA\ (observed in first-order), 
and Fe\,{\sc xiii}
256.43\,\AA, respectively. 
However, we note that our identification of the 175.48\,\AA\ line is the first time this Fe\,{\sc ix}/{\sc x}
feature
has been detected in SERTS data sets, although it has been observed in spectra from other solar
missions (see, for example, Behring et al. 1976). The discrepancy between the 
observed and theoretical
175.48/177.24 ratios indicates that
Fe\,{\sc x} makes about a 30 per cent contribution to the 175.48\,\AA\ line intensity,
in agreement with {\sc chianti}
which predicts 35 per cent. 

The good agreement between theory and observation for the 193.72/177.24 ratio confirms the tentative
identification of the 193.72\,\AA\ line by Del Zanna et al. (2004) in the solar spectrum of Behring et al.
(1976). 
Our detection is the first time this Fe\,{\sc x}
transition has been found in the SERTS data sets. Although we find no evidence of blending,
{\sc chianti} indicates that nearby lines of Ar\,{\sc xii} and Fe\,{\sc xiv}
should contribute about 25 per cent 
to the total
measured 193.72\,\AA\ line intensity. 
Similarly, both the present and {\sc chianti}
theoretical 220.26/177.24 ratios agree with experiment within the errors, 
supporting the tentative identification
of the 220.26\,\AA\ line by Del Zanna et al. in the Behring et al. spectrum.
Once again, this is the first time the
line has been found in SERTS observations.
The measured 220.26/177.24 ratio is approximately 50 per cent
larger than the present theoretical value, 
implying the presence of blending, as also indicated by the large 
line width. This is 
supported by the {\sc chianti}
synthetic spectrum which lists several weak lines of Fe\,{\sc xii}, Fe\,{\sc xv} and Ni\,{\sc xii}
that are predicted to contribute
about 20 per cent 
to the total 220.26\,\AA\ intensity, although as noted above such estimates
must be treated with caution. 

For the 3s$^{2}$3p$^{5}$ $^{2}$P$_{1/2}$--3s$^{2}$3p$^{4}$($^{1}$S)3d $^{2}$D$_{3/2}$  transition in 
Fe\,{\sc x}, we find no emission feature in the SERTS--95 spectrum at the wavelength of
201.49\,\AA\ predicted by Del Zanna et al. (2004). These authors list an alternative
wavelength of 201.56\,\AA, which is based on an identification by Bromage, Cowan \& Fawcett (1977)
from a laboratory plasma. However, the feature at 201.56\,\AA\ in the SERTS--95 spectrum
is clearly not due to Fe\,{\sc x}, as shown by the large discrepancies between theory and observation for 
the 201.56/195.32 and 201.56/177.24 ratios in Table 6. According to the {\sc chianti}
synthetic spectrum, the line is a blend of Fe\,{\sc xi} 201.55 and 201.58\,\AA. 
Although it has previously 
been observed in the solar
spectrum (see, for example, Behring, Cohen \& Feldman 1972), this is the first time the
 201.56\,\AA\ feature has been detected in SERTS data sets.

The excellent agreement between theory and observation for the 195.32/177.24 ratio
confirms our identification
of the 195.32\,\AA\ line of Fe\,{\sc x}, the first time (to our knowledge) that this feature has been observed in
an astronomical source. However, it has been detected in a laboratory spectrum by Bromage et al.
(1977). The experimental 337.24/345.74 ratio is consistent 
with both the present and {\sc chianti}
theoretical values, indicating that the 337.24\,\AA\ line is reliably detected, which is also supported
by the {\sc chianti}
synthetic spectrum where no blending features are listed.
We note that this line was identified as Ar\,{\sc viii}
by Thomas \& Neupert (1994), but is clearly due to Fe\,{\sc x}, as listed by Del Zanna et al.
(2004).

Unfortunately, the 175.27\,\AA\ line is predicted to be N$_{e}$--sensitive 
when its ratio is taken against any other second-order transition of Fe\,{\sc x}
detected in the SERTS--95 spectrum. Hence 
it is not possible to generate 
ratios which are predicted to be independent, or even nearly independent, of the adopted
density.
However, N$_{e}$--diagnostics generated using the 175.27\,\AA\ feature yield 
results consistent
with other ratios (see Section 4.1), and we therefore 
believe the line is free of problems and 
significant blending. We note that the {\sc chianti} synthetic spectrum indicates no likely blending
species, apart from an Fe\,{\sc ix} transition which is only predicted to contribute
about 3 per cent to the total line intensity. 
 
\subsection{Electron density diagnostics}

In Tables 7 and 8 we summarise the observed values of electron-density-sensitive 
line-intensity ratios for the SERTS--89 and SERTS--95 data sets, respectively, 
along with the derived log N$_{e}$
estimates. Only Fe\,{\sc x}
features which are believed to be free of 
significant blends, based on the above assessment,
are considered.
The electron densities in the tables 
have been determined from the present line ratio calculations at the
temperature of maximum Fe\,{\sc x} fractional abundance in ionization equilibrium, T$_{e}$
= 10$^{6}$\,K (Bryans et al. 2006). However, we note that changing T$_{e}$ by $\pm$0.2 dex
would lead to a variation in the derived
values of N$_{e}$ of typically $\pm$0.1 dex or less.
Also listed in the tables is the factor by which the relevant ratio changes 
between N$_{e}$ = 10$^{8}$
and 10$^{11}$\,cm$^{-3}$, to show which are the most N$_{e}$--sensitive diagnostics. 

\begin{table}
 \centering
\begin{minipage}{80mm}
  \caption{Electron density diagnostic line ratios from the SERTS 1989 active region 
spectrum.}
  \begin{tabular}{cccc}
  \hline
Line ratio &   Observed & log N$_{e}$$^{a}$ & Ratio variation$^{b}$ 
\\
\hline
257.25/324.73 & 14 $\pm$ 5 & 9.7$^{-0.2}_{+0.4}$ & 22
\\
257.25/345.74 & 1.8 $\pm$ 0.5 & 9.6$^{-0.4}_{+0.5}$ & 3.5
\\
257.25/365.57 & 3.1 $\pm$ 0.8 & 10.1$^{-0.3}_{+0.5}$ & 3.5
\\
324.73/337.24 & 0.74 $\pm$ 0.33 & 9.2$^{+0.4}_{-0.4}$ & 6.6
\\
324.73/345.74 & 0.13 $\pm$ 0.04 & 9.8$^{+\infty}_{-0.5}$ & 6.2
\\
324.73/365.57 & 0.23 $\pm$ 0.07 & 9.4$^{+0.3}_{-0.3}$ & 6.2
\\
\hline
\end{tabular}

$^{a}$Determined from present line ratio calculations at T$_{e}$ = 10$^{6}$\,K; N$_{e}$ in cm$^{-3}$. 
\\
$^{b}$Factor by which the theoretical line 
ratio varies between N$_{e}$ = 10$^{8}$ and 10$^{11}$\,cm$^{-3}$.
\end{minipage} 
\end{table}

\begin{table}
 \centering
\begin{minipage}{80mm}
  \caption{Electron density diagnostic line ratios from the SERTS 1995 active region 
spectrum.}
  \begin{tabular}{cccc}
  \hline
Line ratio &   Observed & log N$_{e}$$^{a}$ & Ratio variation$^{b}$ 
\\
\hline
175.27/174.53 & 0.20 $\pm$ 0.05 & 9.4$^{+0.2}_{-0.2}$ & 13
\\
175.27/177.24 & 0.40 $\pm$ 0.10 & 9.5$^{+0.2}_{-0.2}$ & 13
\\
175.27/184.53 & 0.82 $\pm$ 0.21 & 9.4$^{+0.2}_{-0.3}$ & 11
\\
190.05/175.27 & 0.36 $\pm$ 0.10 & 9.4$^{-0.2}_{+0.4}$ & 11
\\
257.25/345.74 & 1.7 $\pm$ 0.9 & 9.7$^{-0.9}_{+1.4}$ & 3.5
\\
257.25/365.57 & 4.5 $\pm$ 2.5 & 9.6$^{-1.6}_{+1.4}$ & 3.5
\\
\hline
\end{tabular}

$^{a}$Determined from present line ratio calculations at T$_{e}$ = 10$^{6}$\,K; N$_{e}$ in cm$^{-3}$. 
\\
$^{b}$Factor by which the theoretical line ratio varies between N$_{e}$ = 10$^{8}$ and 10$^{11}$\,cm$^{-3}$.
\end{minipage} 
\end{table}

For the Fe\,{\sc x} transitions reliably
detected in first-order in the SERTS--89 spectrum (i.e.
those between 257--366\,\AA), the best N$_{e}$--diagnostic is 
257.25/345.74, as the lines are strong and unblended.
However, the transitions are far apart in wavelength, so that the ratio is more 
susceptible to possible errors in 
the instrument
intensity calibration.
Also, the ratio is not particularly density sensitive, only changing by a factor of 3.5 
between N$_{e}$ = 10$^{8}$ and 10$^{11}$\,cm$^{-3}$.
If the 
324.73\,\AA\ line is reliably measured (which is difficult as the feature is 
intrinsically weak), then the 324.73/345.74 ratio provides a superior
diagnostic to 257.25/345.74, as the lines are closer in wavelength and 
it varies by a larger factor (6.2) over the 
10$^{8}$--10$^{11}$\,cm$^{-3}$
electron density interval.  
The 257.25/324.73 ratio is even more N$_{e}$--sensitive, 
varying by a factor of 22 between N$_{e}$ = 10$^{8}$ and 10$^{11}$\,cm$^{-3}$,
but has the 
disadvantage of employing lines far apart in wavelength, 
albeit not to the same extent as 257.25/345.74.
We note that the average electron density for the SERTS--89 active region 
from Table 7 is log N$_{e}$ = 9.6$\pm$0.3,
similar to that derived from line ratios in species formed at temperatures
close to that of Fe\,{\sc x}.
For example, from Si\,{\sc x} line ratios (formed at T$_{e}$ = 10$^{6.1}$\,K) Keenan 
et al. (2000) derived log N$_{e}$ = 9.2$\pm$0.2, while from 
S\,{\sc xii}
(T$_{e}$ = 10$^{6.3}$\,K) Keenan et al. (2002) found  log N$_{e}$ = 9.5$\pm$0.3.

\begin{table*}
\centering
\begin{minipage}{140mm}
  \caption{Electron density diagnostic line ratios from EUVE spectra.}
  \begin{tabular}{lccccccc}
  \hline
EUVE data set &   Observed & Observed & log N$_{e}$ & log N$_{e}$ & log N$_{e}$ & log N$_{e}$
& log N$_{e}$
\\
& R$_{1}$$^{a}$ & R$_{2}$$^{a}$ & R$_{1}$$^{b}$ & R$_{2}$$^{b}$ & R$_{1}$$^{c}$ & R$_{2}$$^{c}$  & (other)$^{d}$ 
\\
\hline
Procyon, 1993 January 11 & 0.16 $\pm$ 0.05 & 0.30 $\pm$ 0.09 & 9.2$^{+0.2}_{-0.2}$ & 9.3$^{+0.2}_{-0.3}$ 
& 8.7$^{+0.3}_{-0.3}$ & 8.6$^{+0.3}_{-0.3}$ & 9.4
\\
Procyon, 1994 March 12 & 0.29 $\pm$ 0.07 & 0.45 $\pm$ 0.12 & 9.7$^{+0.3}_{-0.2}$ & 9.6$^{+0.3}_{-0.3}$ 
& 9.3$^{+0.2}_{-0.2}$ & 9.2$^{+0.3}_{-0.3}$ & 9.5
\\
$\alpha$~Cen, 1993 May 29 & 0.15 $\pm$ 0.04 & 0.20 $\pm$ 0.06 & 9.2$^{+0.2}_{-0.2}$ & 9.0$^{+0.2}_{-0.3}$ 
& L$^{e}$ & L & 9.2
\\
\hline
\end{tabular}

$^{a}$R$_{1}$ = 175.27/174.53; R$_{2}$ = 175.27/177.24; from Foster et al. (1996).
\\
$^{b}$Determined from present line ratio calculations at T$_{e}$ = 10$^{6}$\,K. 
\\
$^{c}$Determined from Fe\,{\sc x}
line ratio calculations of Foster et al. (1996) at T$_{e}$ = 10$^{6}$\,K.
\\
$^{d}$Determined from other line ratios by Foster et al. (1996).
\\
$^{e}$Indicates that the experimental line ratio is smaller than the theoretical low density limit.
\end{minipage} 
\end{table*}

An inspection of Tables 7 and 8 clearly reveals that the best N$_{e}$--diagnostics
for Fe\,{\sc x} lie in the second-order wavelength 
range (175--190\,\AA) 
covered by the SERTS--95 active region spectrum. In particular, 
the 175.27/174.53 and 175.27/177.24 ratios involve transitions which are strong and unblended plus
close in wavelength. Additionally, both ratios vary by large factors (13) between
N$_{e}$ = 10$^{8}$ and 10$^{11}$\,cm$^{-3}$.
The 175.27/184.53 and 190.05/175.27 ratios also potentially provide good diagnostics, 
although the relevant transitions
are somewhat further apart in wavelength and hence more susceptible to 
possible errors
in the instrument intensity calibration.
We note that the values of N$_{e}$ deduced for the SERTS--95 active region from the 4 ratios above
are all consistent,
with an average of log N$_{e}$ = 9.4$\pm$0.1. This is in excellent agreement
with the value of log N$_{e}$ = 9.4$\pm$0.2 derived by Brosius et al. (1998b)
from Fe ions formed at similar temperatures to Fe\,{\sc x}.

Previously, Foster et al. (1996) calculated the 175.27/174.53 and 175.27/177.24 
ratios of Fe\,{\sc x}
using the 
electron impact excitation rates of Mohan, Hibbert \& Kingston
(1994), and employed these as N$_{e}$--diagnostics
for Procyon and 
$\alpha$~Cen via a comparison with observations from the Extreme Ultraviolet Explorer (EUVE) satellite.
However, Foster et al. found that the electron densities derived from Fe\,{\sc x}
showed large discrepancies (up to an order of magnitude) with those deduced from 
other density-sensitive line ratios in the EUVE spectra.
To investigate this, in Table 9 we list the EUVE line ratio measurements
for Fe\,{\sc x}
along with the electron densities derived from the Foster et al. and present line ratio
calculations.
An inspection of the table 
reveals that the electron densities deduced from the current theoretical estimates of 175.27/174.53
and 175.27/177.24 are consistent, and furthermore agree 
with the values of N$_{e}$
determined from other line ratios, hence resolving the discrepancies found by Foster et al.
However, we should point out 
that the discrepancies are also resolved if theoretical line ratios from
{\sc chianti} are employed.

Finally, we note that Datla, Blaha \& Kunze (1975) have measured several Fe\,{\sc x}
extreme-ultraviolet line ratios in a $\theta$-pinch under well-determined plasma conditions,
hence allowing some independent 
assessment of the accuracy of our theoretical results.
In Table 10 we summarise their experimental line ratios,
along with the theoretical values for the measured plasma conditions from both the present calculations
and {\sc chianti}.
An inspection of the table reveals that the measured 190.05/174.53 ratios are always larger than theory,
which is expected as the 190.05\,\AA\ line will be blended with Fe\,{\sc xii}
190.07\,\AA\ in the $\theta$-pinch spectra, as these are of somewhat lower resolution
(approximately 0.1\,\AA) than the SERTS data, where the two features are resolved (see above).
For the other ratios, agreement between theory and experiment is 
reasonable, given that the measurements are estimated to be accurate to only 20--30 per cent.
In most instances, agreement is slightly better with the present line ratio calculations,
but once uncertainties are taken into account the observations are consistent with both sets of 
theoretical results.
The Datla et al. measurements therefore
provide some experimental support for the two atomic physics data sets
adopted in the relevant line ratio calculations. 

\begin{table*}
\centering
\begin{minipage}{140mm}
  \caption{Comparison of theoretical line ratios with $\theta$-pinch plasma measurements.}
  \begin{tabular}{lcccc}
  \hline
Measured plasma parameters &  Ratio & Observed$^{a}$ & Present & {\sc chianti} 
\\
& & & theory$^{b}$ & theory
\\
\hline
 T$_{e}$ = 1.65$\times$10$^{6}$\,K; N$_{e}$ = 1.2$\times$10$^{16}$\,cm$^{-3}$
& 177.24/174.53 & 0.81 & 0.57 & 0.56
\\
& 190.05/174.53 & 0.22 & 0.087 & 0.076 
\\
& 365.57/174.53 & 0.032 & 0.031 & 0.021 
\\
 T$_{e}$ = 1.10$\times$10$^{6}$\,K; N$_{e}$ = 1.4$\times$10$^{16}$\,cm$^{-3}$
& 177.24/174.53 & 0.59 & 0.58 & 0.55
\\
& 190.05/174.53 & 0.14 & 0.088 & 0.077 
\\
& 365.57/174.53 & 0.021 & 0.040 & 0.025
\\
\hline
\end{tabular}

$^{a}$From Datla et al. (1975).
\\
$^{b}$Present theoretical line ratios and those from {\sc chianti} calculated for the measured plasma parameters. 
\end{minipage} 
\end{table*}

\section{Conclusions}

Our comparison of theoretical Fe\,{\sc x}
emission-line intensity
ratios with solar active region spectra from the SERTS 1989 and 1995 flights
reveals generally
very good agreement between theory and experiment, with several features identified
for the first time in the SERTS data sets, including Fe\,{\sc x}
193.72, 220.26 and 324.73\,\AA, plus Fe\,{\sc xi} 201.56\,\AA\ and the
Fe\,{\sc ix}/{\sc x} blend at 175.48\,\AA. 
In addition, the 195.32\,\AA\ transition of Fe\,{\sc x}
is detected for the first time (to our knowledge) in an astronomical source.

We find that the ratios 175.27/174.53 and 175.27/177.24 provide the best Fe\,{\sc x}
electron density diagnostics, as they involve lines which are strong and free from
blends, are close in wavelength and the ratios 
are highly N$_{e}$--sensitive. Should these lines not be available, then the 
257.25/345.74 ratio may be employed as a diagnostic, although this requires an accurate
determination of the instrument 
intensity calibration over a relatively large wavelength range.
However, if the weak 324.73\,\AA\ line is reliably detected, then the use of 
324.73/345.74 or 257.25/324.73 is recommended in preference to 257.25/345.74.

\section*{Acknowledgments}

KMA acknowledges financial support from EPSRC, while
DBJ is grateful to the Department of Education and Learning
(Northern Ireland) and NASA's Goddard Space Flight Center
for the award of a studentship.
The SERTS rocket programme is
supported by RTOP grants from the Solar Physics Office   
of NASA's Space Physics Division.                        
JWB acknowledges additional NASA support under
grant NAG5--13321.                                       
FPK is grateful to AWE Aldermaston for the award of a William Penney
Fellowship. The authors thank Peter van Hoof for the use of his
Atomic Line List. {\sc chianti} is a collaborative project 
involving the Naval Research Laboratory (USA), Rutherford
Appleton Laboratory (UK), and the Universities of Florence 
(Italy) and Cambridge (UK).

\bsp

\label{lastpage}

\end{document}